\documentclass[twocolumn,showpacs,showkeys,preprintnumbers,superscriptaddress,amsmath,floatfix,amssymb,secnumarabic,nofootinbib]{revtex4-1}
\usepackage[colorlinks=true]{hyperref}
\usepackage{graphicx}
\usepackage{dblfloatfix} 
\usepackage{placeins}
\usepackage{braket}
\usepackage{amsmath}
\usepackage{epsfig}
\usepackage{float}
\usepackage{nicefrac}
\usepackage{xcolor}
\usepackage{color}
\usepackage{bm}
\usepackage{subfigure}
\usepackage{chngcntr}
\usepackage{ulem}

\def\({\left(}
\def\){\right)}
\def\[{\left[}
\def\]{\right]}

\newcommand{\ve}[1]{\boldsymbol{#1}}

\newcommand{\beq} {\begin{eqnarray}}
\newcommand{\eeq} {\end{eqnarray}}

\newcommand{\comment}[1]{ \textcolor{cyan}{ #1} }

\begin{document}
\sloppy

\title{ 
Validity of SLAC fermions for the (1 + 1)-dimensional helical Luttinger liquid}

\author{Zhenjiu Wang}
\email{zhwang@pks.mpg.de}
\affiliation{Max-Planck-Institut f\"ur Physik komplexer Systeme, Dresden 01187, Germany}

\author{Fakher~Assaad}
\email{Fakher.Assaad@physik.uni-wuerzburg.de}
\affiliation{Institut f\"ur Theoretische Physik und Astrophysik, Universit\"at W\"urzburg, 97074 W\"urzburg, Germany}
\affiliation{W\"urzburg-Dresden Cluster of Excellence ct.qmat, Am Hubland, 97074 W\"urzburg, Germany}

\author{Maksim~Ulybyshev}
\email{Maksim.Ulybyshev@physik.uni-wuerzburg.de}
\affiliation{Institut f\"ur Theoretische Physik und Astrophysik, Universit\"at W\"urzburg, 97074 W\"urzburg, Germany}

\begin{abstract}
The Nielson-Ninomiya theorem states  that  a chirally invariant free fermion lattice action, which is local, translation invariant, and real necessarily has fermion doubling. The  SLAC  approach   gives  up on locality  and long  range hopping leads  to a   linear  dispersion  with  singularity at the zone boundary. 
We introduce a SLAC Hamiltonian formulation that is  expected  to realize a U(1) helical Luttinger liquid in a naive continuum limit. 
	We  argue  that  non-locality and concomitant  singularity  at the zone  edge   has  important implications.  Large momentum  transfers  yield  spurious  features already in the non-interacting  case.  Upon  switching on interactions  non-locality  invalidates the Mermin-Wagner  theorem and allows  for long ranged  magnetic ordering.   In  fact,  in the  strong  coupling  limit the  model  maps  onto an  XXZ-spin  chain  with  $1/r^2$ exchange.  Here, both spin-wave   and  DMRG  calculations  support  long  ranged order.      
While  the long-ranged  order  opens  a single particle  gap  the  Dirac point, the  singularity at  the  zone-boundary  persists for  any  finite 
value  of   the interaction  strength  such  that  the  ground state  remains metallic.   
Hence,  SLAC Hamiltonian   does  not  flow   to  the $1$d helical Luttinger liquid  fixed point. 
Aside  from  DMRG  simulations, we   have  used  auxiliary  field   quantum Monte  Carlo   simulations   to arrive  to the  above  conclusions. 
\end{abstract}

\maketitle

\section{\label{sec:Intro}Introduction}

The Nielson-Ninomiya theorem states  that  a chirally invariant free fermion lattice action, which is local, translation invariant, and real necessarily has fermion doubling  \cite{Nielsen81}.  How should one then carry out simulations of  a single Dirac cone?  A possible route is to consider higher dimensions. A single Dirac cone  in say 1+1-dimensions  can be realized as a  surface state of a 2+1-dimensional topological insulator.    The other Dirac cone lies on the  other surface and as the system size grows mixing between the  cones will  vanish  such that the physics of a  single cone can be studied.  In the realm of  high-energy physics, this construction  is referred to as domain wall fermions  \cite{KaplanB92}.    In the  domain of the  solid  state,  this  idea  has   been  used to  study  correlation effects in   helical  Luttinger liquids \cite{Hohenadler10,Hohenadler11a}. 

 Alternatively  one can violate  one of the assumptions of the  Nielsen-Ninomiya   theorem.  SLAC fermions   are subject to long range
  hopping and thereby  violate the locality  condition.    They have  been  used in a  number of  solid  state \cite{Lang18,Li18,Tabatabaei22,Daliao22} and high energy physics \cite{PhysRevD.96.094504,PhysRevD.100.054501,PhysRevD.101.094512}  setups, and  seem to  provide  a simple  route to simulate  a  single  Dirac  cone  in a lattice model   with  finite 
  lattice  constant  $a$.     In  particular,   it  avoids   the    potentially  expensive  step  of  dealing  with  higher  dimensional systems. 
  SLAC   fermions   come  with  a singularity  at  the     Brillouin  zone  boundary  at   $k = \pm \pi/a $  in one   dimension. 
The  question we will ask  in this  article  is how   the  non-locality  and   concomitant    singularity   at  the zone  edge    effects   the    physical  results, 
in comparison  to a   domain wall  fermion  approach.  
  
 To  do so, we will  consider  the simplest possible  model,  the helical   Luttinger liquid   emerging  at the  boundary of  a 2D quantum spin Hall  insulator 
 as  realized  by  the Kane-Mele  model \cite{KaneMele05}.    In particular  we will consider  a  setup   with  U(1)   symmetry,  corresponding to  conservation of   z-total  spin.      This    choice is   challenging  for SLAC fermions.    For  short ranged  interactions the  Mermin-Wagner  theorem  states that  continuous  symmetries   cannot  be  spontaneously broken in 1+1-dimension even in the  zero  temperature limit.   In fact, in  conjunction with   the  intrinsic  nesting instabilities of 1+1-dimensional  systems  this  impossibility of  ordering  leads   to the  fluctuation 
 dominated  physics  of the  Luttinger  liquid \cite{Haldane81,Giamarchi}.       The non-locality  of  SLAC  fermions violate  the assumptions of the
 Mermin-Wagner  theorem  and can hence lead  to  artifacts  especially  in the strong  coupling limit.    We note that this has  recently been 
 pointed out in Ref.~\cite{Daliao22}.

Another reason for the choice of this model is that  its naive continuum limit,   corresponding  to  ignoring the  zone boundary singularity, can be solved exactly  since only   forward  scattering   is  allowed.   The  results  of  the  bosonization approach   have  been  favorably  compared  to  
 calculations based on  domain wall  fermions  \cite{Hohenadler10,Hohenadler11a}.  
 In this  article we  formulate  a SLAC  Hamiltonian 
 that allows  for  negative  sign 
 free  auxiliary field  quantum Monte Carlo (QMC)   simulations.  
 The key question that we want to ask is that, is it meaningful to compare   
 between the physics of SLAC Hamiltonian and the one of Luttinger liquid?

 The  article is organized as  follows.  In the next  section, \ref{sec:generalities},   we will discuss  the   SLAC   formulation  of the  helical Luttinger  liquid.     Before  discussing  our  results  for  the non-interacting and  interacting cases in Sec.~\ref{sec:results},   we  will summarize  the   bosonization results  in Sec.~\ref{sec:bosonization}   and    the  technicalities  of  the Monte  Carlo  simulations in  Sec.~\ref{sec:qmc}.      
 In Sec.~\ref{sec:strong_coupling}    we  discuss  a simple  model to understand  our  strong coupling results.  In Sec.~\ref{sec:conclusions}   we  summarize  the implications of  our  results.   
The  article   contains several  Appendices   that   demonstrate  the absence  of  negative sign problem   (see  App.~\ref{appendixA}), that  discuss  the   scaling dimension of  $\hat{S}^z_i$   as  a  function of the coupling  strength (see App.~\ref{appendixB}),  and  that  provide  a  spin wave  analysis  of the     long  ranged  XXZ  model  (see App.~\ref{appendixC}).  

\section{\label{sec:generalities}  SLAC  formulation of   the helical  liquid  }
We   consider the following one-dimensional  model of length $L$  and  lattice  constant $a$:  
\begin{eqnarray}
\label{Slac1d.eq} 
\hat{H} =  & &   -v_F \sum_{i=1} ^{L} \sum_{r=-L/2}^{L/2} t(r) \left(  \hat{a}_{i}^{\dagger} \hat{b}_{i+r}^{} + \hat{b}^{\dagger}_{i+r}  \hat{a}_{i}^{}   \right)   \nonumber  \\ 
 & & +  U \sum_{i}  \left( \hat{n}^{a}_{i} - \frac{1}{2} \right)  \left( \hat{n}^{b}_{i} - \frac{1}{2} \right) 
\end{eqnarray}
with 
\begin{equation}
\label{Slac_t.eq}
t(r)  = (-1)^{r}  \frac{\pi}{L \sin \left( r \pi/L\right) }   \text{  for  }  r \neq 0  \text{  and  }   t(0) = 0.
\end{equation}
Each unit cell harbors two orbitals,  and $\hat{a}_{i}^{\dagger} $,  $\hat{b}_{i}^{\dagger} $,   are spinless fermion  creation operators.

\begin{figure}[htp]
\begin{center}
\includegraphics[width=0.5\textwidth]{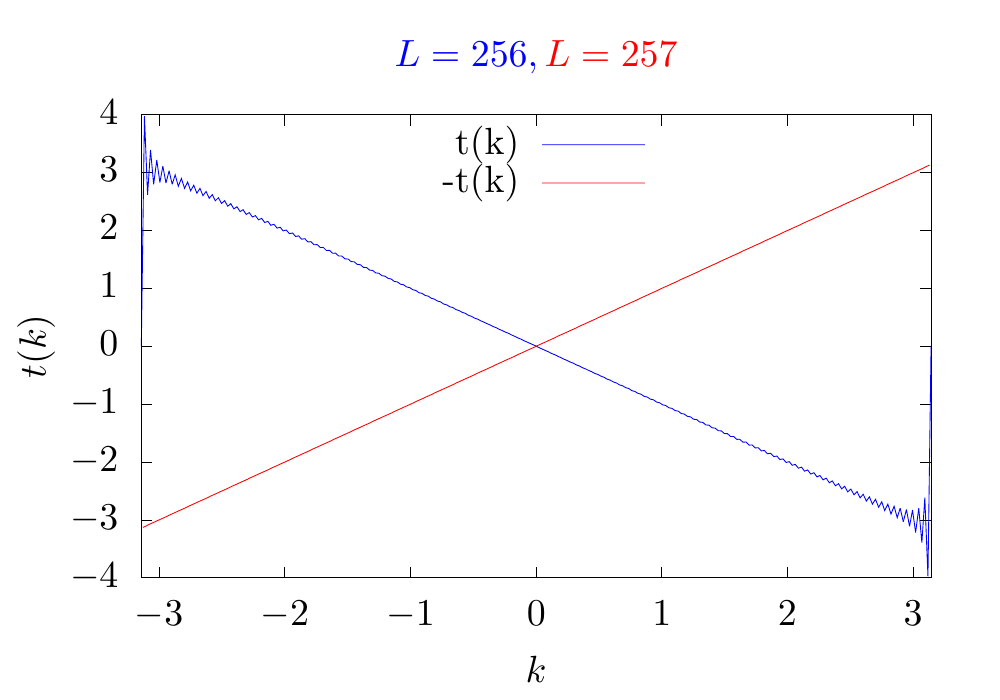}
\end{center}
\caption{ $t(k)$    for even and odd lattices. The Gibbs phenomenon is apparent on even lattices.     Here  we  set $a=1$ }\label{SLAC.fig}
\end{figure}

Using  periodic  boundary  conditions and Fourier transformation,
\begin{equation}
  \begin{pmatrix}
  \hat{a}^{}_{k} \\
  \hat{b}^{}_{k}    
  \end{pmatrix}   =  \frac{1}{\sqrt{N}}   \sum_{j=1}^{L}   e^{i k j   } 	
   \begin{pmatrix}
  \hat{a}^{}_{j} \\
  \hat{b}^{}_{j}    
  \end{pmatrix}
\end{equation}
gives,  up to a  constant, 
\begin{eqnarray}
  \hat{H} =  & &  -v_F \sum_{k=-\frac{\pi}{a}}^{\frac{\pi}{a}}   t(k) \left( \hat{a}^{\dagger}_{k},     \hat{b}^{\dagger}_{k} \right)   \boldsymbol{\sigma}^{y}   
  \begin{pmatrix}
  \hat{a}^{}_{k} \\
  \hat{b}^{}_{k}    
  \end{pmatrix}    \nonumber  \\ 
   & &   - \frac{U}{2} \sum_{i}  
  \left[ \left( \hat{a}^{\dagger}_{i},     \hat{b}^{\dagger}_{i} \right)   \boldsymbol{\sigma}^{x}    
  \begin{pmatrix}
  \hat{a}^{}_{i} \\
  \hat{b}^{}_{i}    
  \end{pmatrix} \right]^{2}
\end{eqnarray}
	with   
\begin{equation}
 t(k) =  i \sum_{r=-L/2}^{L/2} e^{-i k r}  t(r).
\end{equation} 
In the  above,    $L = N a $. 
For any lattice size, $t(k)$ is  a real and odd function. It is plotted in Fig.~\ref{SLAC.fig}  and as apparent  scales to  $t(k) =k$   for  $k$ in the  f Brillouin zone (BZ)  and  in the
 thermodynamic limit. One will also notice the  Gibbs phenomenon (on even lattices) at the zone boundary associated to the discontinuity of $t(k)$. 
	
The rotation: 
\begin{equation}
\label{eq:rot}
 \begin{pmatrix}
  \hat{a}^{}_{i} \\
  \hat{b}^{}_{i}    
 \end{pmatrix}   = \frac{1}{\sqrt{2}}  \left(  1 + \boldsymbol{\sigma}^{x}      \right) 
  \begin{pmatrix}
  \hat{c}_{i,\uparrow}\\
  \hat{c}_{i,\downarrow}    
 \end{pmatrix} 
\end{equation}
gives
\begin{equation}
\label{SLAC_HL.eq}
     \hat{H} = -v_F  \sum_{k \in BZ, \sigma}  \sigma   k   \hat{c}^{\dagger}_{k,\sigma} \hat{c}^{}_{k,\sigma}    +  U \sum_{i}  \left( \hat{n}_{i,\uparrow} - 1/2 \right) 
     \left( \hat{n}_{i,\downarrow} - 1/2 \right) 
\end{equation}
corresponding to a helical liquid with Hubbard interaction. Our   goal  is to investigate  if  the   
SLAC approach indeed reproduces the  expected results  obtained from bosonization of the helical liquid. 

\section{ Results  from bosonization }
\label{sec:bosonization}
 Let us  start  by  stating  the  bosonization   \cite{Giamarchi,Hohenadler11a} results  valid in the  continuum limit,  $a \rightarrow 0$.
In this  limit,    the  fermion field  operator  reads: 
 \begin{equation}
    \hat{\Psi}_{\sigma}(x)    = 
      e^{i k_f x } \hat{R}(x) \delta_{\sigma,\uparrow}  +  e^{-i k_f x } \hat{L}(x) \delta_{\sigma,\downarrow}   
 \end{equation}
 where  $\hat{R}(x)$  and   $\hat{L}(x) $   are   independent  right  and  left  propagating  fermion operators  with  spin and    direction of  motion  locked in.  Inserting  the  above   form in Eq.~\ref{SLAC_HL.eq} gives: 
 \begin{eqnarray}
 	\hat{H} =  & & - v_F  \sum_{k}   k \left( \hat{R}^{\dagger}_{k}    \hat{R}^{}_{k} - \hat{L}^{\dagger}_{k}    \hat{L}^{}_{k}  \right)    +   \nonumber  \\
	   & &  Ua  \int_{0}^{L} \, d x \,   \hat{R}^{\dagger}(x)    \hat{R}^{}(x)   \hat{L}^{\dagger}(x)    \hat{L}^{}(x) 
 \end{eqnarray}
 where $ \hat{R}_k  =  \frac{1}{L} \int_0^{L}   dx  \, e^{i k  \, x }\hat{R}(x) $.    For  a  short ranged model  with   nearest neighbor  hopping matrix element $t$,     $ v_F  = 2 t a $.     Hence,  to obtain a  well  defined  continuum limit,      we  scale  both   $t$  and  $U$  as   $1/a$.      Since  we have  taken the  continuum limit,   the   sum  over  momenta is  unbounded. 
 The  above   forward  scattering model  can  be  solved   with  bosonization 
 techniques   reviewed  in Ref.~\cite{Giamarchi}. Correlation  functions 
are   given by:   
\begin{eqnarray}
\label{correlations.eq}
	& & C_{n}(r)  \equiv \langle  \hat{n}(r)  \hat{n}(0) \rangle   \propto   \frac{1}{r^2}  \nonumber   \\
	& &C_{S^z}(r)   \equiv  \langle  \hat{S}^z(r)  \hat{S}^z(0) \rangle   \propto   \frac{1}{r^2}  \nonumber   \\
	& & C_{S^x}(r)   \equiv  \langle  \hat{S}^x(r)  \hat{S}^x(0) \rangle   \propto   \frac{\cos(2 k_f r ) }{r^{2K_{\rho}}} \\ 
	& & C_{\Delta}(r)  \equiv  \langle  \text{Re} \hat{\Delta}^{\dagger}_r  \text{Re} \hat{\Delta}^{\phantom\dagger}_0  \rangle   \propto \frac{1}{r^{2/K_{\rho}}} \nonumber
\end{eqnarray}
In the  above  $K_{\rho} $ is  a  interaction strength  dependent   Luttinger liquid  exponent, $\hat{n}(r)  =  \sum_{\sigma} \hat{c}^{\dagger}_{r,\sigma}  \hat{c}_{r,\sigma} $, $\ve{S}(r)   =  \frac{1}{2} \ve{c}^{\dagger}_r  \ve{\sigma} \ve{c}^{\phantom\dagger}_r  $ with  $\ve{\sigma}$  a vector  of Pauli 
spin matrices  and   $   \hat{\Delta}^{\dagger}_r  =  \hat{c}^{\dagger}_{r,\uparrow}  \hat{c}^{\dagger}_{r,\downarrow} $.     
Here,  $2k_f$ denotes  the  momentum  difference   of the left  spin up  and  right spin down  movers. This  wave  vector  is  naturally  picked up  
in  $C_{S^x}(r)$  since  it involves  scattering  between the  two branches.   In our   construction,  $k_f=0$.     Before  proceeding and  
as  mentioned  earlier  the  bosonization results  are  consistent  with  the  domain   wall fermion  approach of  \cite{Hohenadler10,Hohenadler11a}  
even in the  rather  strong  coupling limit. 
The  above  will  be  our  reference   result  and  we will ask the question  under   which conditions we  can reproduce it with the  SLAC  lattice  regularization. 

\section{Quantum Monte Carlo simulations}
\label{sec:qmc}

The Hamiltonian of Eq.~\ref{Slac1d.eq}   does not suffer from a negative sign 
problem in QMC  simulations \cite{Li16,Wei16}.    To see  this,   one   will rewrite  the model  as, 
\begin{eqnarray}
\label{Slac1d_QMC.eq} 
\hat{H} =  & &   -v_F \sum_{i=1} ^{L} \sum_{r=-L/2}^{L/2} t(r) \left(  \hat{a}_{i}^{\dagger} \hat{b}_{i+r}^{} + \hat{b}^{\dagger}_{i+r}  \hat{a}_{i}^{}   \right)   \nonumber  \\ 
 & & - \frac{U}{2} \sum_{i}  \left( \hat{a}^{\dagger}_{i} \hat{b}^{\phantom\dagger}_{i}  + \hat{b}^{\dagger}_{i} \hat{a}^{\phantom\dagger}_{i} \right)^2 
\end{eqnarray}	 
where we  have omitted  a  constant.  
Next  we  adopt  a  Majorana  representation,
\begin{equation}
\label{majorana:eq}
	\hat{a}_i = \frac{1}{2} \left( \hat{\gamma}_{i,1,1} + i \hat{\gamma}_{i,2,1} \right),   \,\,  \,  \hat{b}_i = -\frac{i}{2} \left( \hat{\gamma}_{i,1,2}  + i \hat{\gamma}_{i,2,2}  \right), 
\end{equation}
to obtain, 
\begin{eqnarray}
\label{Slac1d_QMC1.eq} 
\hat{H} =  & &   v_F \sum_{i=1} ^{L} \sum_{r=-L/2}^{L/2} t(r)   \frac{i}{4}\hat{\ve{\gamma}}_{i}^{T} \tau_x  \hat{\ve{\gamma}}_{i+r}^{}     \nonumber  \\ 
 & & - \frac{U}{2} \sum_{i}  \left( \frac{1}{4}\hat{\ve{\gamma}}_{i}^{T} \tau_y \hat{\ve{\gamma}}_{i}^{}   \right)^2, 
\end{eqnarray}	    
where $\hat{\ve{\gamma}}_{i}^{T} = \left( \hat{\gamma}_{i,1,1},  \hat{\gamma}_{i,2,1},    \hat{\gamma}_{i,1,2},  \hat{\gamma}_{i,2,2} \right)  $.   Here  we  
have used  the fact  that  $t(r)$ is an  odd  function of  $r$  and adopted  the  notation   
$ \hat{\gamma}_{i,\sigma, \tau} $   where  the  Pauli  $\tau$   ($\sigma$)-matrices act  of the $\tau$   ($\sigma$)   
indices.   A  global  O(2)   symmetry  in the  $\sigma$-indices  now becomes  apparent. 
After  a  real   Hubbard-Stratonovich  transformation of  the  perfect  square,    the  Fermion  determinant  will  be  given by  the  square of  a Pfaffian.   Since one  will show  that the  Pfaffian is real,    we will  conclude in the absence of the negative sign problem.    Hence  the  absence of  sign problem  for this  SLAC  model of  the helical  liquid  follows    the  same  logic  as for the so  called  spin-less  t-V model  \cite{Yao14a,Huffman14}.       In the appendix    we  show   the  absence  of  sign problem    for  the  generic  model: 
 \begin{eqnarray}
\label{Slac1d_QMC1.eq} 
\hat{H} =  & &   -v_F \sum_{i=1} ^{L} \sum_{r=-L/2}^{L/2} t(r) \left(  \hat{a}_{i}^{\dagger} \hat{b}_{i+r}^{} + \hat{b}^{\dagger}_{i+r}  \hat{a}_{i}^{}   \right)   \nonumber  \\ 
 & &  - \frac{U}{2}  \sum_{i} \left[
       \left( \hat{a}^{\dagger}_{i}, \hat{b}^{\dagger}_{i} \right)   \sigma_\alpha  \left( \hat{b}^{\phantom\dagger}_{i}, \hat{a}^{\phantom\dagger}_{i} \right)^{T}
      \right]^2
\end{eqnarray}	 
where   $\sigma_{\alpha}$ is  a Pauli spin  matrix.     Note that  after    computing the square  one  will explicitly see  that the  Hamiltonian is  $\alpha$-independent.   For   any  value  of   $\alpha$,  we can  use   the  ALF  \cite{ALF_v2}  implementation of the  finite temperature  auxiliary  field  QMC  algorithm  \cite{Blankenbecler81,Hirsch85,White89,Assaad08_rev}.     In fact,   Eq.~\ref{Slac1d_QMC1.eq}    that  formulates  the  interaction in terms of a  perfect  square,   has   the  required  form  for  usage  of the  ALF-library,  and  concomitant  Hubbard-Stratonovich  transformation. 

As mentioned above,  the  results  are  $\alpha$-independent.   However  the  Monte  Carlo Markov  chain  will have a  strong  $\alpha$-dependence.    
We  have  seen  that   we obtain the best  results  when  considering the  $\sigma_{y} $   formulation.   This stems form  the  fact  that  after  
the rotation of  Eq.~\ref{eq:rot}     the  U(1)   symmetry   of  the helical  liquid is  satisfied  for  each  Hubbard-Stratonovich  field   configuration.

We also would like to stress that  since  we  are working in the Hamltonian formulation,    the   resulting  Lagrangian  has   SLAC hoppings  only  in the  spatial direction  and  is  local  along   the Euclidean time direction.

We  used the interaction  strength (band width)  
as the energy unit for simulations at large (small) values of $U/v_F$.    
For $U / v_F \leq 4 $ we   choose $ v_F \beta  = L $ and $ v_F \Delta \tau  = 0.1$;   
whereas for $U / v_F > 4$ we considered  $ U \beta / 4 =  L $ and  $ U \Delta \tau /4 = 0.1 $.

\section{Results}
\label{sec:results}

We will show  that  the SLAC approach   suffers   from two basic issues.

The first one can be seen already in the non interacting limit and  originates  from processes with large momentum transfer. This deficiency can be illustrated as the violation of the anomaly relation in lattice Schwinger model with SLAC fermions \cite{NASON1985269}. If we consider the continuum theory and turn on  a constant electrical field pointing to the right, the right movers will acquire momentum and energy and will fill their branch of the dispersion relation up to some positive level. At the same time the left movers will loose  energy  and hence their branch of the cone will be filled only up to the same but negative level. Thus the axial charge will appear as an imbalance between the right and left movers. However, this is not true for the SLAC fermions due to the finite size of the Brillouin zone and finite depth of the Dirac sea.  
E.g. the right movers at the bottom of the Dirac sea will also acquire energy thus the very bottom of this branch of the dispersion relation will not be filled any more. These effects will compensate the difference between right- and left-movers in the low momentum modes leading to the axial charge being always zero. Though this is only a  qualitative illustration, it shows the presence of the non trivial dynamics near the edge of the Brillouin zone.

Another issue can be seen upon  switching on 
intermediate to  strong correlations  as  measured  in  unit of the  band-width.  In this case we observe a long ranged order 
again contradicting the results from continuum theory. 

Both points will be carefully studied in the subsequent sub-sections. 

\subsection{The non-interacting  case}
SLAC  fermions  become   very  transparent when introducing a length  scale   $\xi$ in the hopping:
\begin{equation}
 t_{\xi}(r)   = t_0(\xi)  \frac{ (-1)^{r/a} \pi}{ \frac{L}{a}  \sin(r \pi /L )}  e^{ -  \sin(r \pi /L )/\xi}.
\end{equation}
In the above,  we  adjust   $t_0(\xi)$  so as  to fix  the bandwidth to $2\pi$.    Clearly, the  Fourier  transform of  a  short range  function 
has  to be smooth  and one  will see  that for  any  finite  value of $\xi$  we   observe  two  crossings  of the Fermi  surface  albeit  with very 
different  values  of the velocity.  In fact  the  velocity  at the zone boundary  diverges  with  growing values of   $\xi$. 
\begin{figure}[htp]
\begin{center}
\includegraphics[width=0.5\textwidth]{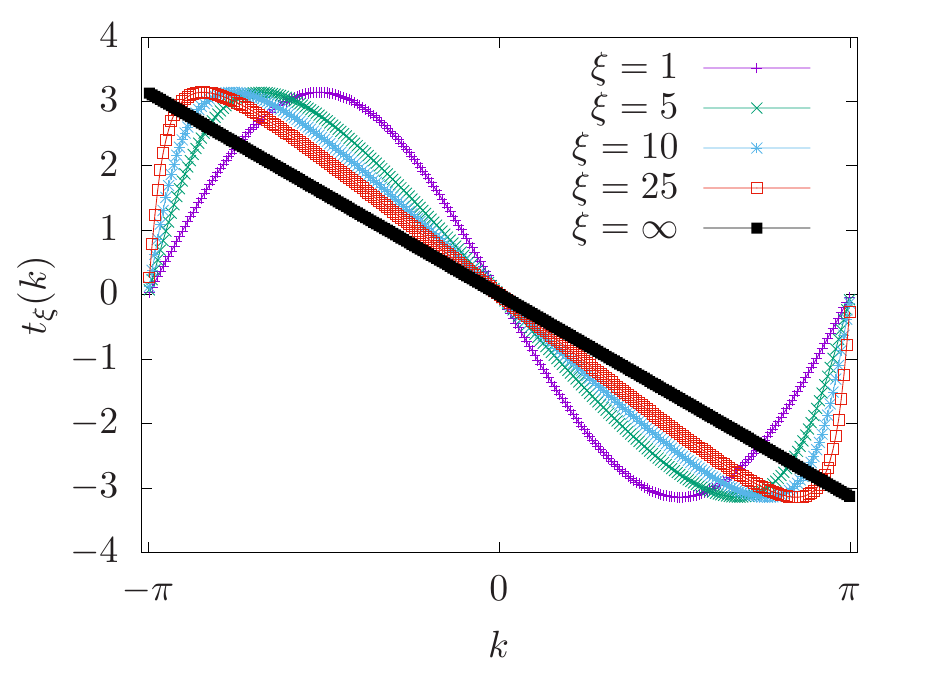}
\end{center}
\caption{ $k$ dependence of $t_{\xi}(k)  $  for length  scale  $\xi=1,5,10,25$ and $\infty$.  
We took $L=257$.  
 }\label{SLAC_1.fig}
\end{figure}
In principle,  for  any  finite value of  $\xi$   we  expect  Umklapp  processes  to be  relevant  such that  any  finite value of  $U$  should  lead  to 
an insulating state.    However,  since the velocity at the zone boundary  diverges  as  $\xi$,  the  phase space  available to  these 
Umklapp process will vanish in the $\xi \rightarrow \infty $  limit.      Another   consequence  of the  singularity  at the  zone boundary  is  that 
\textit{large}   momentum transfer   will always  provide  a   discrepancy  with the  bosonization even in the non-interacting case.  One 
can illustrate  this by considering the  charge-charge  correlation functions  for SLAC Hamiltonian for the  half-filled  case, $\mu=0$,  at  zero  temperature:  
\begin{equation}
	\langle \hat{n}(r)  \hat{n}(0) \rangle =  \frac{1}{2 \pi^2}   \frac{  \cos( \pi r) - 1 } {r^2}.
\end{equation}
In  the  above $\hat{n}(r)  =  \sum_{\sigma} \hat{c}^{\dagger}_{r,\sigma}  \hat{c}_{r,\sigma} $.  This result is  independent on the   value of  $\xi$  and  merely  relies on the fact  that the  dispersion relation    intersects  the  Fermi  energy  at  wave vectors  $k = 0$ and $k = \frac{\pi}{a} $.   The above expression already deviates from the bosonization result \ref{correlations.eq} and shows  that  already  at this  level one  will obtain  the   same  result  as  for the continuum model,  where  the  zone  edge  diverges, only if one  blocks  large momentum transfers.  This can be done  by  introducing  point-splitting operators on the lattice, as was already suggested in \cite{PhysRevD.26.839,PhysRevD.101.094512}. 

\subsection{Monte Carlo  results}

\begin{figure*}
\centering
\includegraphics[width=\textwidth, height=0.96\textheight ]{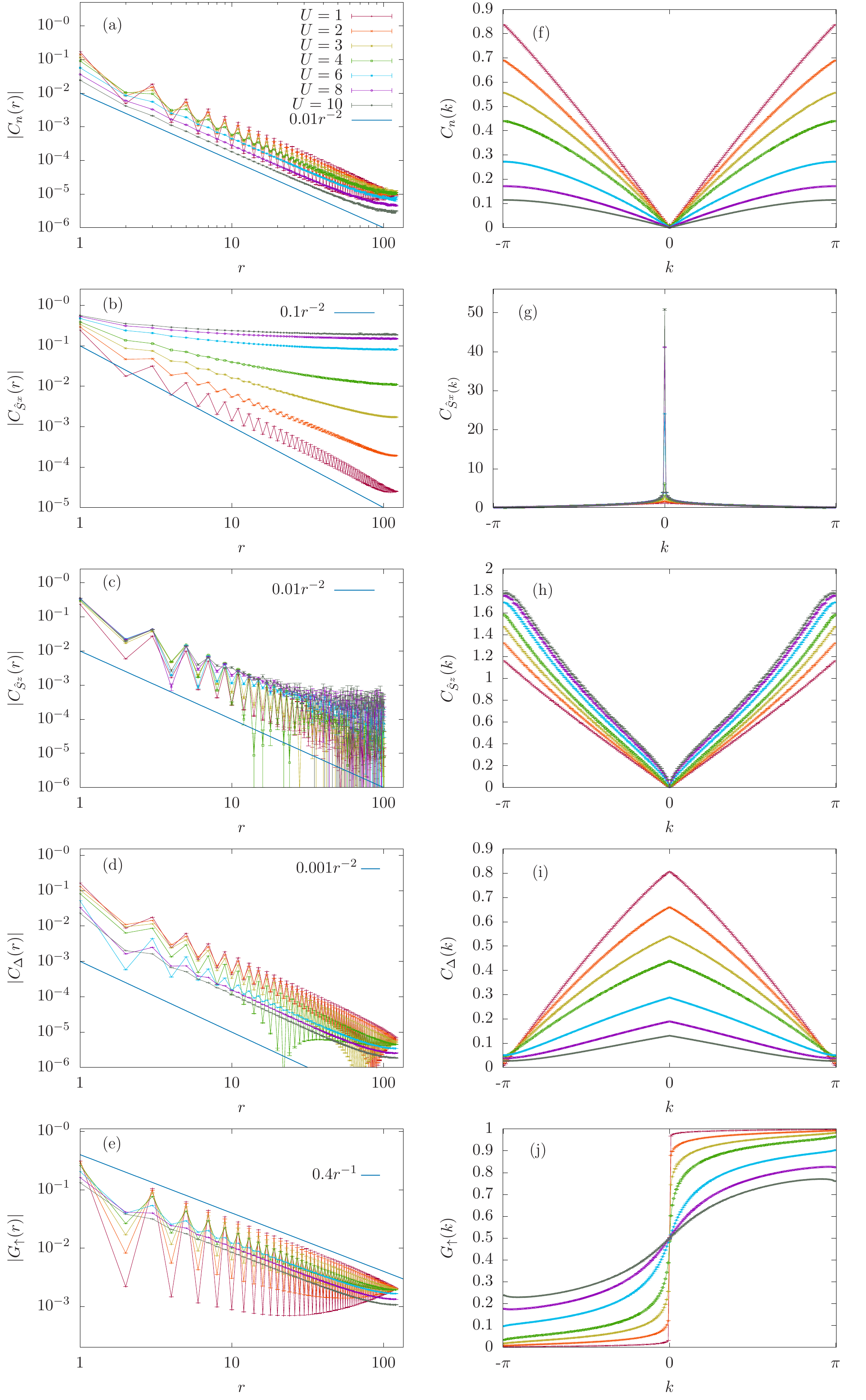}    
\caption{   
 Real space  correlation   functions  (a)-(e)    and  corresponding   structure  factors   (f)-(j).   Here  we  consider the  charge, $C_n$, x (z)-component of  spin  $C_{S^x}$ /$C_{S^z}$ ),    pairing  $C_{\Delta}$  and  single  particle  $G_{\uparrow}$   correlation  functions.    All the subfigures share the same 
  legend color as the one in (a).    
 For $C_n$, $C_{S^x}$, $C_{\Delta}$ and $G_{\uparrow}$, we    chose $L=243$;  whereas for $C_{S^z}$, we considered  
 $L=203$.  The reason of this mismatch are large fluctuations in the  QMC runs for  $C_{S^z}$ and for large sizes. 
}\label{fig:C} 
\end{figure*}

\begin{figure}
\centering
\includegraphics[width=0.45\textwidth,height=0.9\textheight]{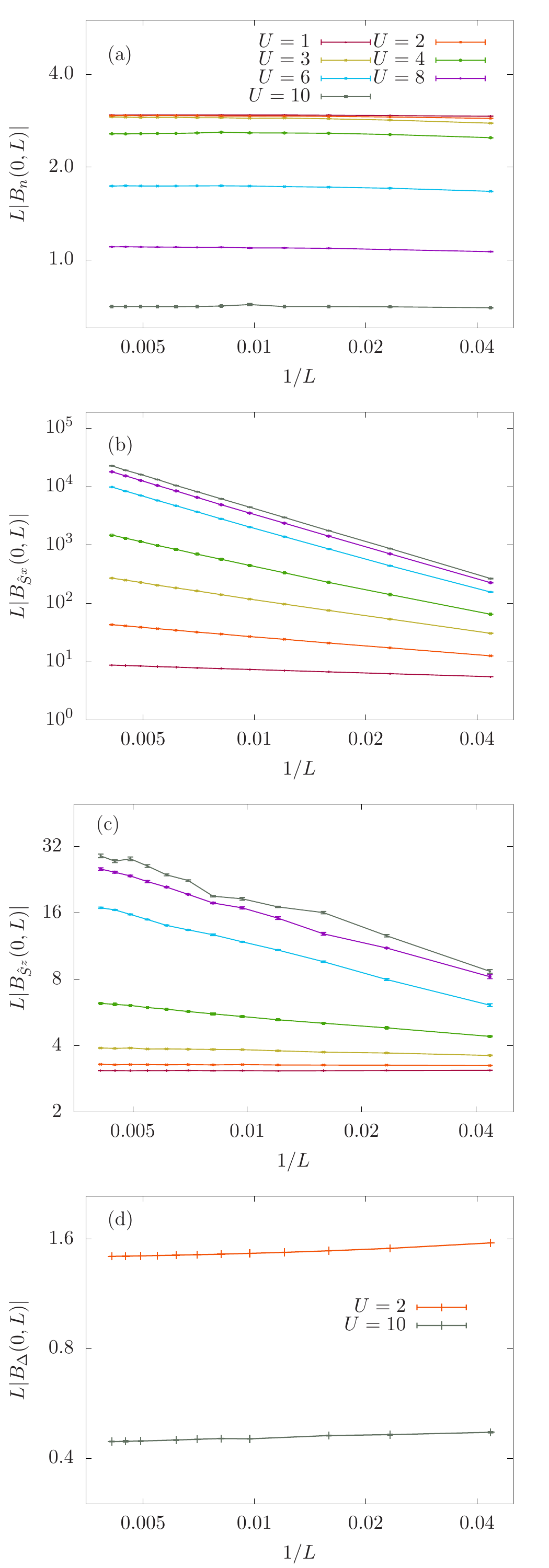}     
\caption{  
  $ B( k = 0, L ) L $ as function of $1/L$ for charge (a), $\hat{S}^x$ (b), $\hat{S}^z$ (c), pairing (d). 
   (b) and (c) share the same 
  legend color as the one in (a).  
}\label{fig:LB} 
\end{figure}

We  have  computed     structure  factors:
\begin{equation}
	S_{\bullet}(k)   =   \sum_{r}e^{i k r }  C_{\bullet}(r) 
\end{equation} 
where  the  bullet  refers  to  charge,  spin  along the z- or x-spin quantization axis or paring  correlations (see Eq.~\ref{correlations.eq}).    
 To  obtain an  estimate 
of the power law    decay  at a given  wave  vector, one  can consider: 
 \begin{eqnarray}
B_{\bullet}(k, L/2) \equiv 2C_{\bullet}(k)  -C_{\bullet}(k+2\pi/L)-C_{\bullet}(k-2\pi/L).  \nonumber  \\
  \label{SF}
\end{eqnarray}
$ \frac{L}{2 \pi}  B_{\bullet}(k, L/2)   $  corresponds  to  the  left  minus   the  right  derivative    at  a  given  $k$-vector.     Hence  for  a   smooth  function this  
quantity  scales  to zero as a function of system size.    However  for  $k$-vectors  where  one observes a  cusp,  it  will scale  to a finite  value. 
One  will show  that: 
\begin{equation}
 C_{\bullet}(L/2)
=\frac{1}{4L}\sum_{k} e^{i k L/2}B_{\bullet}(k, L/2)
\end{equation}
such  that the scaling of $B_{\bullet}(k, L/2)$  at  wave  vectors k   where one observes a cusp will  reflect  the  
decay of  the correlation function \cite{Assaad91}  at  this  wave  vector.

Fig.~\ref{fig:C}   plots  the  real   and  k-space  correlation functions   for  the  above mentioned  quantities.
  To better   understand  the  results,   we  consider  the 
behavior  of  the cusps in the  corresponding   structure  factors  by  plotting  $ B_{\bullet}(k=0, L/2)  $   as  a function
of  system size and   coupling  strength. 

Let  us  start with the  charge.  From  Fig.~\ref{fig:C}(a)   we  see  that  irrespective of  the  coupling constant in the  range  $U 
\in \left[ 0, 10 \right] $   the   real  space   charge  correlation  decays  as  $1/r^2$.    In the  weak  coupling limit  we  observe    a  $(-1)^{r}$ modulation  
alongside  the  uniform   decay.   This  weak  coupling behavior   gives  way to a uniform  decay  at  strong coupling.    In  k-space,   Fig.~\ref{fig:C}(f)  we   see  that  the cusp  at  $k=\pi$  rounds  off as a function of   growing interaction strength  but that the  cusp at  $q=0$  remains  robust.   We  also notice 
that  as  a function of growing  interaction  strength  the  charge  response  is  suppressed.   We  can pin  down  the charge  exponent  by  analyzing   
 $ L B_{n}(k=0, L/2) $  in  Fig.~\ref{fig:LB}(a).   Irrespective of  the interaction strength,  it  is  to an  accurate  degree  $L$-independent    thus   reflecting   a   $1/r^2$  decay  of  the  charge correlations.

 At  weak  coupling the z-component of  spin  is  very   similar  to the  charge   (at   $U=0$ they  are  identical),  see  Figs.~\ref{fig:C}(c,h).    In  contrast  however,  the    cusp at  $k=0$ becomes  more  pronounced at  strong coupling.   Fig.~\ref{fig:LB}(c),    shows  that  the z-spin  correlations  acquire  a
 non-trivial  exponent   in the  strong  coupling limit.  This stands at odds  with the bosonization result  of  Eq.~\ref{correlations.eq}.

The  spin-correlations  along the  x-spin quantization, Figs.~\ref{fig:C}(b,g),  direction  are  most  intriguing  results.      At  weak  coupling  and on our  system  sizes,  this  correlation  function  follows    roughly  a $1/r^2$   form  consistent  with  $LB_{S^x}(k=0,L) $  constant, Fig.~\ref{fig:LB}(b).    
$L B_{S^x}(k=0, L/2) $    has  a  marked  U-dependence.   It  is  remarkable  to  see  that  at    strong  coupling  $L B_{S^x}(k=0, L/2)    \propto  L^2 $     thus  suggesting  long  range  magnetic order as is confirmed  by  
the very  strong  peak  in the  structure  factor  a  $k=0$ and  the  lack of  decay  in real space.    We note that  $L B_{S^x}(k=0, L/2)    \propto  L^2 $  is   consistent  with  $C_{S^x}(k=0)    \propto  L $.   This result  seems at  odds with the  Mermin-Wagner  theorem,  that  states  that a  
continuous  symmetry  cannot   be  broken  at $T=0$  in the  ground state.  However,  the assumptions  for  the theorem to be  valid   requires  \textit{short ranged}  interactions.   The  non-locality  of  the SLAC  fermions  may very  well  invalidate   this assumption.    
We  note  that  long  range  order  can be  stabilized  by  coupling  spin-chains  locally  to an ohmic  bath  thus introducing long  ranged  interactions  along  the imaginary time \cite{Weber22}. 

Long  ranged order   along  the x-spin quantization axis   breaks  time reversal  symmetry  and allows  for  elastic  scattering between the   right-moving  spin down and  left  moving spin  up  electrons.   At  the  single particle,  mean-field   level  we  expect: 
 \begin{equation}
 \label{MF.eq}
 	  \hat{H}_{MF}     =  \sum_{k=-\frac{\pi}{a}}^{\frac{\pi}{a}}  \hat{\ve{c}}^{\dagger}_{k}  \left[- v_f k \sigma_z    +   \comment{U}m_x \sigma_x \right]  \hat{\ve{c}}^{\phantom\dagger}_{k}  
 \end{equation} 
 where  $m_x$  denotes  the  ordered  moment.     This symmetry  breaking   generates a  mass gap  at  the    Fermi  momentum   $k_f=0$:
  $E_k =  \pm \sqrt{ \left(v_f k\right)^2  +   \left(\comment{U}m_x\right)^{2}} $.      Fig.~\ref{fig:C}(j),   plots  the  single particle equal time  Green  function,
  $G_{\sigma}(k) =  \langle  \hat{c}^{\dagger}_{k,\sigma} \hat{c}^{\phantom\dagger}_{k,\sigma}  \rangle $.   As  $U$  grows, the   singularity  at  $k=0$   evolves to a  smooth   feature.    Fig.~\ref{fig:C}(e)  confirms  this:  at  weak  coupling, $G_{\sigma}(r)   \propto 1/r$  as  expected for    Dirac  electrons  in 1+1D,  and in the  strong  coupling limit  the   $L$-independent  from of  
  $L  B(k=0,L) $  is   consistent  with a mass  gap.    $G_{\sigma}(k) $  has  another  singularity  at  $k=\pi$     that  dominates  the 
  long-ranged real  space  behavior.   Putting all together,   the  data in the  strong coupling limit   is  consistent  with the form: 
  $G_{\sigma}(r)    \propto   \left( a e^{-r/\xi} +  (-1)^{r} \right) /r$  where  $\xi$  is   set  by  the inverse   ordered moment  $m_x$.  
  We also  notice  that  the overall  amplitude of  $G_{\sigma}(r) $    diminishes  as  a   function of  U  in the  strong coupling. 
  
  Finally,  we  consider  the  pairing correlations in  Figs.~\ref{fig:C}(d,i) as  well as  in Fig.~\ref{fig:LB}(d).  We again observe  non-analyticities at 
  $k=0$ and  $k=\pi$ in the   structure factor.  The non-analytical  behavior  at  $k=0$ survives the  strong   coupling limit,  whereas $ C_{\Delta} (k ) $  evolves  towards a smooth function in the  vicinity of  $k=\pi$.   The  singularity  at $k=0$ leads  to a  $1/r^2$  decay  of   the  pair correlation,  and again the overall  amplitude  of  the correlation function  decreases as  as function of increasing $U$.

 \section{Interpretation of  the strong  coupling  limit}
\label{sec:strong_coupling}
In this section,  we  provide a consistent  interpretation of the   strong coupling limit.   In this limit  the QMC  data  shows   long ranged 
magnetic ordering   along  the  x-spin quantization axis.   The   single  particle  Green  function  decays  as   $(-1)^r/r$,    and  
the   density  as  well as  the  pairing   correlations    follow a  $1/r^2$  law.    On the  other  hand  the  z-component of  spin  correlations  
 have  a power-law decay   with  exponent depending   on the  interaction strength.  Clearly   this behavior  lies at odds   with the 
bosonization  results.

 Our    simulations  shows  that   the  ground is  a   metal   with  long ranged    magnetic  order.       To at  best   understand  our  results,    let us  start  with a mean-field    representation of this strong  coupling 
ground state:  
 \begin{equation} 
     | \Psi_0 \rangle   =   \prod_{r}   \frac{1}{\sqrt{2}} \left(  \hat{c}^{\dagger}_{r,\uparrow} +  \hat{c}^{\dagger}_{r,\downarrow} \right) | 0 \rangle. 
 \end{equation}
Since the above  wave  function has  no charge fluctuations  and precisely  one  electron per  site  it is  the    ground state
of  the  Hubbard interaction  term, 
$\hat{H}_U$  with  energy  $E_0$.    
 \footnote{Here  we omit    spin fluctuations    discussed  at  length in  App.~\ref{appendixB}  and \ref{appendixC}}  
Consider  the  small hopping  limit  such  that the  ground state  wave  function  can be  estimated  
 perturbatively 
in   the  hopping  $\hat{H}_t$
\cite{Messiah}:
\begin{equation}
\label{eq:second_order_pert}
	 | \Psi \rangle     =     | \Psi_0  \rangle  + \hat{Q}_0  \frac{1}{\hat{H}_U -  E_0} \hat{Q}_0 \hat{H}_t | \Psi_0  \rangle.
\end{equation}
In the  above  $\hat{Q}_0  = 1 - | \Psi_0  \rangle \langle \Psi_0 | $.  Let  us  now  compute the  charge  fluctuations, 
$C_{n}(r)= \langle  \Psi |  \left( \hat{n}_r -1\right)  \left(\hat{n}_0 - 1 \right)  |  \Psi  \rangle  $  for $r \neq 0$.   The  sole  contribution reads: 
\begin{eqnarray}
	C_{n}(r)  = & &  \langle   \Psi_0  |    \hat{H}_t    \hat{Q}_0  \frac{1}{\hat{H}_U -  E_0} \hat{Q}_0     \left( \hat{n}_r -1\right)  \left(\hat{n}_0 - 1 \right)    \times         \nonumber   \\ 
	  & &     \quad  \hat{Q}_0  \frac{1}{\hat{H}_U -  E_0} \hat{Q}_0 \hat{H}_t  | \Psi_0  \rangle.
\end{eqnarray}
Since  $\hat{H}_t$   has  hopping processes on all   length scales   it    contains the operator   $ t(r) \sum_{\sigma}   \sigma \hat{c}^{\dagger}_{r,\sigma} 
\hat{c}^{\phantom\dagger}_{0,\sigma} $.  Applied on  $ | \Psi_0  \rangle $   it  will generate  a  doublon-holon   pair at  distance  $r$  with an  energy cost
with respect  to  $E_0$  set by $U$.   This charge  fluctuation  will be picked up by   $  \left( \hat{n}_r -1\right)  \left(\hat{n}_0 - 1 \right)  $.  Finally  the 
doublon-holon  pair   will be  destroyed  by  the operator $ t(r) \sum_{\sigma}   \sigma \hat{c}^{\dagger}_{0,\sigma} 
\hat{c}^{\phantom\dagger}_{r,\sigma}  $  again  contained  in $\hat{H}_t$.   As a result,  we  estimate: 
\begin{equation}
	C_{n}(r)    \simeq   \frac{t^2(r)}{U^2}  \propto   \frac{1}{U^2    r^2}.
\end{equation}
The  power-law  is  confirmed  by the  QMC  data  of  Fig.~\ref{fig:C}(a).  It is also interesting to note  that the magnitude    of the  charge-charge  correlations      are predicted  to scale  as   $1/U^2$.    Comparison  between  the  $U=6$ and  $U=10$  data in Fig.~\ref{fig:C}(f)   supports  this scaling. 

  The  very same  argument can  be carried out  for the pairing correlations.   Let us pick up the above argument  at  the point where  
doublon is  created on site $r$  and  a  holon on site  $0$.  Applying the  pairing  operator $ \hat{\Delta}_{r}  \hat{\Delta}^{\dagger}_0 $    on this  state,     will yield  a  non-zero result  and  transfer   the doublon  (holon)  to the  origin (site r).      The operator  $ t(r) \sum_{\sigma}   \sigma \hat{c}^{\dagger}_{r,\sigma}  \hat{c}^{\phantom\dagger}_{0,\sigma} $  will then  annihilate the  doublon-holon pair,  and we  will obtain a  finite  overlap  with the  mean-field 
ground state.   Hence  we  also  expect: 
\begin{equation}
	C_{\Delta}(r)    \simeq   \frac{t^2(r)}{U^2}  \propto   \frac{1}{U^2    r^2}  \label{eq:C_delta_der}
\end{equation}
in the  strong coupling limit, which is consistent with our  QMC  data, but inconsistent with the bosonization results Eq.~\ref{correlations.eq}.

We now    consider  the  single particle Green  function.    Here,    the relevant  terms  in 
$  \langle \Psi  | \hat{c}^{\dagger}_{r,\sigma} \hat{c}^{\phantom\dagger}_{0,\sigma}  |  \Psi \rangle  $      are  the  mixed  terms of the form: 
\begin{equation} 
	   \langle \Psi_0 |  \hat{c}^{\dagger}_{r,\sigma} \hat{c}^{\phantom\dagger}_{0,\sigma}   \hat{Q}_0  \frac{1}{\hat{H}_U -  E_0} \hat{Q}_0 \hat{H}_t | \Psi_0  \rangle.
\end{equation}
The  doublon-holon  pair    created  by  $\hat{H}_t$  will   be annihilated   by   the   single particle   transfer   $\hat{c}^{\dagger}_{r,\sigma} \hat{c}^{\phantom\dagger}_{0,\sigma}  $.     In accordance  with the QMC   results  this approximation  gives: 
\begin{equation}
	G_{\sigma}(r)    \propto  \frac{t(r)}{U}     \propto  \frac{(-1)^{r}}{U r}. 
\end{equation}

We  now comment  on the  nature, metallic  or insulating,   of  the  strong  coupling   wave  function.    The  very  fact  that the  charge
 correlations  follow  a
 power-law,     suggest  a metallic  ground state.     An  accepted  definition of  an insulating or  metallic  state  is  the  Drude  weight \cite{Kohn64}, that  
 probes  the  localization  of   the  wave  function.    Here,  one   considers  a   ring  geometry   and  threads  a magnetic  flux  $\Phi$  through the  
 the  center of the   ring.   Such a flux  will have  an effect  if    the  charge  carriers   are  delocalized  and  can  circle around it and, owing to the Aharonov-Bohm   effect,   acquire   a  phase  factor    $e^{2\pi i \Phi/\Phi_0}$   where  $\Phi_0 $ is the flux  quanta.   Here  we  assume  that the  charge  carriers  have  the electron charge. The  Drude  weight   in  d-spatial  dimensions is  defined  as: 
 \begin{equation}
 	  D(L) =  \frac{1}{L^{d-2}} \left.   \frac{\partial^2 E_0(\Phi)}{\partial \Phi^2}  \right|_{\Phi=0}. 
 \end{equation} 
 For  the  insulating state   $D(L)$    vanishes exponentially  with   $L$  reflecting the localization  length  of the  wave function.    For    a   metallic  state  the  Drude  weight is  finite.   Let us  now  use   this accepted  criterion  to the  SLAC fermions, in the  strong coupling limit.     A  glimpse  at the  wave  function in second order perturbation theory  (see Eq.~\ref{eq:second_order_pert})   shows  that it  contains   holon-doublon  excitations,  at  all  length  scales.   The  fact  that they  are  costly in energy, means  that they  are  short lived,  but  during  this short  time,  they can  propagate  over large  distances  due  to the non-locality  of the   hopping.   Hence  we   expect the  Drude  weight to be finite.    To  substantiate this  statement we  carry out  the following estimations.   The   flux  leads  to   a  twist in the  boundary  condition: 
 \begin{equation}
 	  \hat{c}_{r+L,\sigma}   = e^{2 \pi i  \frac{\Phi}{\Phi_0}}  \hat{c}_{r,\sigma}   
 \end{equation}
 that    we  can   rid  of   with  the canonical   transformation: 
  \begin{equation}
 	  \hat{d}_{r,\sigma}   = e^{-2 \pi i  \frac{\Phi}{\Phi_0} \frac{r}{L}}  \hat{c}_{r,\sigma}. 
 \end{equation}
 Under  this  canonical     transformation,   the  Hubbard  term   remains  invariant,  the hopping  reads,
 \begin{equation}
 	\hat{H}_t(\Phi)  =   v_F i \sum_{i,\sigma}  \sum_{r}   \sigma t(r)    \hat{d}^{\dagger}_{i,\sigma}\hat{d}^{\phantom\dagger}_{i+r,\sigma} e^{2 \pi i  \frac{\Phi}{\Phi_0} \frac{r}{L}},
\end{equation}
 and  $  \hat{d}_{r,\sigma} $   satisfies  periodic  boundary   conditions: 
  $  \hat{d}_{r,\sigma}   =  \hat{d}_{r+L,\sigma} $. 
   Let  us  now  compute  the  second  order  contribution  to  the  energy  that will  pick  up the    dependence  on the flux: 
\begin{equation}
   	E_2(\Phi)  =  \langle  \Psi_0  |  \hat{H}_t(\Phi) \hat{Q}_0  \frac{1}{\hat{H}_U -  E_0} \hat{Q}_0 \hat{H}_t(\Phi) | \Psi_0  \rangle
\end{equation}
Starting  from  $| \Psi_0  \rangle $   one  will create   for example a  holon at  position $i$   and   a  doublon at  position $i+r$   by  applying  the hopping.  
 This    process   has   matrix  element    $i v_F  t(r)  e^{-2 \pi i  \frac{\Phi}{\Phi_0} \frac{r}{L}}$  and  energy  cost    set  by  $U$.    
 The only  way  to perceive  the  flux  is  for  the  charge excitation to encircle it.  Hence  the  second hopping process  should    destroy  the  doublon in favor  of  single occupancy at  site $i+r$   and  create  an  electron  at  site $i+L \equiv  i$ thereby   restoring  single  occupancy  on this site such that the overlap  with $|\Psi_0 \rangle $  does not  vanish. This  second  process   comes  with matrix  element:  $ i v_F  t(L-r)  e^{-2 \pi i  \frac{\Phi}{\Phi_0} \frac{L-r}{L}} $.   Putting  everything  together  one obtains: 
 \begin{equation}
 	E_2(\Phi)  \propto  - \frac{v_F^2}{U} \sum_i \sum_{r}   t(r)  t(L-r)   \cos\left( 2 \pi  \frac{\Phi}{\Phi_0} \right).
 \end{equation}
 We  hence  see  that   in this  approximation,  the  Drude  weight   reads: 
 \begin{equation}
 	  D   \propto  L^{2}  \left(\frac{2 \pi}{\Phi_0}\right)^2  \frac{v_F^2}{U} \sum_{r}   t(r)  t(L-r).  
 \end{equation}
 One  will  check that $\sum_{r}   t(r)  t(L-r)$   takes a  finite  value.  Hence   we  obtain  the   result  that the Drude weight  actually  diverges as  $L^2$,  and  only at  $U= \infty$  will we  have  an   insulating state on any  finite lattice.

The   above   real  space  picture    does not provide an    explanation  of  the observed  power-law  decay  of  the 
 spin correlations  along the z-quantization axis.   Our  perturbative  calculation   creates  a  doublon-holon   pair,  and   since  these excitations   carry  no  spin,      $C_{S^z}(r)$     vanishes  identically.    To  go  beyond  this approximation,   we  can   consider  the  Hamiltonian of  Eq.~\ref{MF.eq}.   In fact  in the  limit  $U \rightarrow  \infty$   this approximation will  reproduce  the   above  perturbative  results.      Given, Eq.~\ref{MF.eq}  we    can   compute   $C_{S^z}(r)$    to  obtain: 
 \begin{equation}
 	C_{S^z}(r)   =  - \frac{1}{2} \left| \frac{1}{L} \sum_{p   =  -\frac{\pi}{a}}^{\frac{\pi}{a}} 
	   e^{-i p r} \frac{v_F p} {\sqrt{ \left(v_F p\right)^2 +  \left(  U m_x  \right)^2    }} \right|^2.    
 \end{equation}
  The  sum under  the  square    corresponds  to the  single particle  Green  function, that,  due  to  the singularity  at  the  Brillouin zone  edge  decays 
  as   $1/r$  with a   $(-1)^{r}$  modulation.    Since  the spin-correlation is a  particle-hole    excitation,  it   decays  as  $1/r^2$    but  with  no  spatial  
  modulation.   Furthermore,     in the  strong  coupling   limit,   the   amplitude  of  the  spin-spin correlations  along the z-quantization axis  would    scale  as  $1/U^2$.   The  above   stands  at  odds  with the  QMC  data.   As shown in Fig.~\ref{fig:LB}, 
$C_{\hat{S}^z}$  seems to pick up a  non-trivial scaling  dimension in the sense that it  decays slower than $1/r^2$, as $U$ increases to a scale comparable to the band width.   Furthermore in the  strong  coupling limit  Fig.~\ref{fig:C}(h)     shows  that   the  amplitude  of     $ C_{S^z}(k)  $   
grows as  a function  of  increasing  $U$.     Hence,  the  data  begs  for  another interpretation.    

As  seen above,  in the  limit $ U \rightarrow \infty $      charge    fluctuations  are     suppressed  by  a   factor  $1/U^2$  such  that  we  can  carry 
out  a  Schrieffer-Wolff  transformation    to  obtain  the  Heisenberg model: 
\begin{equation}
\label{Strong.eq}
     \hat{H}_{U  \rightarrow  \infty}  =  \frac{4v_f^2}{U} \sum_{i,r=-L/2}^{L/2}   t^2(r) \left[ \hat{S}^{z}_{i}  \hat{S}^{z}_{i +r}   -   \hat{S}^{x}_{i}  \hat{S}^{x}_{i +r}    - \hat{S}^{y}_{i}  \hat{S}^{y}_{i +r}    \right].
\end{equation} 
Since $t^2(r)  \propto  1/r^2 $  the   conditions for the  validity  of   Mermin-Wagner  theorem  \cite{Auerbach}  are   not  satisfied.    Furthermore,   the  spin interaction along  the z-direction is  antiferromagnetic,   thus  leading  to frustration     due  to the long  ranged  nature of the   exchange.   Since  in the  transverse  direction  the  coupling is   ferromagnetic,  frustration can be avoided by  ordering in the   x-y  plane. 
In fact, spontaneous  $U(1)$ symmetry breaking of this spin model has been confirmed by numerical and renormalization group analysis \cite{Gorshkov_XXZ},   and naturally  the   magnetic  ordering is reproduced by  our  simulations at large $U$ limit.

Furthermore, in appendix.~\ref{appendixB}, we systemically show extrapolation of 
 $\hat{\Delta}_{S^z}$ 
as function of $U$: $\Delta_{ \hat{S}^z}$ starts to deviate from $1$ at intermediate ranges of $U$ and approaches around $0.7$ at large $U$ limit.  Hence fluctuations around   the    mean-field  approach have to be taken into consideration.  At this point, we only have solid understanding for the scaling behavior of the XXZ chain in the large $U$ limit.  In   appendices  \ref{appendixB}   and   \ref{appendixC}    we    carry  out   density matrix renormalization group  simulations  and  linear  spin-wave  calculations, to  show that  the   scaling   dimensions  of  $\hat{S}^{z}_{i}$,  $\Delta_{\hat{S}^{z}}  = 3/4$.  As  a  consequence,    the  spin  structure  factor  $C_{S^z}(k)   \propto \sqrt{k}$   in the long  wave-length limit.   
 This is consistent with the decay of scaling dimension in SLAC system as strength of the interaction  grows.

\section{\label{sec:conclusions}   Discussion  and  Conclusions}

We introduce a one dimensional toy model Hamiltonian based on SLAC fermion approach.  
Our   SLAC model   differs  from the  1+1  dimensional  Helical   liquid  by  a  singularity  at the   Brillouin-zone  boundary.   
Although we originally aimed at benchmarking the validity of this approach in describing  the 1+1  dimensional  Helical   liquid, a completely different fixed point is found.  
For  this  very  specific  case,   we   understand  that  the  differences  are  present  both in  the  
weak  and  strong  coupling  limits.   Hence  the singularity at  the  zone  boundary  is    a  relevant  perturbation  at  the  1+1 dimensional Helical liquid 
fixed  point.

One  dimensional  systems  are   generically nested.   For  the   helical  Luttinger  liquid  at  $U=0$  of   Eq.~\ref{SLAC_HL.eq}, this leads  to  $  \chi_{\perp}(k=0,\omega=0)  \propto   \log \frac{v_f}{k_B T}  $.  As a  consequence,  a   mean-field approach  to  correlation effects  will  generate    long-ranged  magnetic  order    along  the spin-x quantization axis  and  a  charge gap.   Both  the  charge gap  and  the ordered moment  will   follow  an  essentially   singularity  in  the weak  coupling limit.     

For  generic   local one-dimensional    models   we  know that  the  above  Stoner  arguments cannot  be made  due  to    the Mermin-Wagner 
 theorem \cite{Auerbach}   that  tells  us  that quantum   fluctuations will  destroy  the  ordering  even in the  ground  state.   
 For  our  specific  case,   continuous   U(1)  spin-symmetry  breaking is   not  allowed.   This competition between   the  Stoner  instability  and  the  Mermin-Wagner   theorem  is  
 at the  very  origin of  Luttinger   liquid  behavior  generic  to 1+1D   interacting  systems.        This is  exemplified   by  the  helical  Luttinger  liquid:  
 a  metallic  state  with no single particle  gap  and  an  interaction  strength dependent 
power-law  decay of  the spin-spin  correlations  in the transverse  direction.      We  note  that  due to U(1) spin-symmetry  Umklapp  processes   are  symmetry     forbidden such  that   the  system  will  remain  metallic  for  arbitrary  large   interactions.  This understanding of the  helical   Luttinger  liquid,  has  been  confirmed  numerically within  
a    domain-wall  fermion  approach \cite{Hohenadler10,Hohenadler11a}   in  which  interaction effects  are  included  only on one  set of  domain-wall  fermions.

The non-locality  of the SLAC  fermion approach   brings  major  differences  to  the  above  picture.   The  key-point   is  that   it   violates the  assumptions of the Mermin-Wagner  theorem.  
The  violation of  the Mermin-Wagner  theorem in the  realm of  SLAC  fermions  was  recently  pointed  out  in  Ref.~\cite{Daliao22}.      
Our  numerical  
results  explicitly  confirm  this   in the  strong coupling limit  where long ranged  magnetic order along the x-spin 
quantization axis  and   global U(1)  spin  symmetry  breaking is observed.    This  allows   for  a mass  term  and  in fact  we  observe  a single  particle  gap opening at  the 
Fermi  wave  vector.

We should note that one can try to use the SLAC fermions in the context, when the continuous chiral symmetry is reduced to a  $Z_2 $ discrete one. For instance, such situation emerges when one considers more than one flavor of fermions in $1+1$D. If the interaction term is written as $(\bar\psi_a \psi_a)^2$, where $\psi$ is two-component spinor, the continuous symmetry is broken and only $Z_2$ symmetry remains. Analogously, the spin-orbit  coupling will    reduce the  U(1)  continuous  symmetry  to a  $Z_2 $    discrete one.  In this  case,  the 
aforementioned  issues of  SLAC action  with the Mermin-Wagner theorem  will  be  waived. 
However,  some artefacts will likely survive even  for  the  discrete $Z_2$ symmetry.  In particularly, the deviation of the behaviour of the correlation functions in Eq. \ref{eq:C_delta_der}    from the   strong  coupling limit does  not involve a continuous  symmetry for its derivation.  Hence these discrepancies will remain even  for models  with  discrete  symmetries.  Another important point is the nature of the ordered state observed in our QMC simulations. In contrast  to   Dirac  systems where   magnetic mass terms   are  generated  spontaneously \cite{Assaad13},  this  ordered  state  remains  metallic. 
This is again a consequence of non-locality  inherent  to SLAC  approach  that   produces    doublon-holon pairs at  any  length  scale.      Equivalently,  the  current  operator  
becomes  long-ranged.   Strictly  speaking  Gross-Neveu    transitions  that  have beed  studied   in the  realm of  SLAC  fermions    \cite{Lang18,Tabatabaei22,Daliao22} 
are  not  metal  to insulator  transitions but  metal-to-metal  ones.  

At  vanishing  coupling  strength,  the   results  of the  Helical-Luttinger  liquid  with  $K_{\rho}=1$    become  exact  provided  that we  block  large  momentum transfers.    As  mentioned  above,  this non-interacting    point is  unstable.       An important question is to  asses  if  there
is  a  finite value of  $U_c$  below  which  we  will observe  the physics  of  the  Helical  Luttinger  liquid.  We  conjecture  that  $U_c=0$. 
As  mentioned  above,    the non-interacting  limit  is  unstable to ordering  in  the  transverse  spin direction.     Since  the non-locality of  the model   leads  to a    violation of the  Mermin-Wagner theorem   quantum fluctuations  will not   destabilise the  ordering  and  will not  invalidate a  Ginzburg-Landau  mean  field     picture.    In this  case,  the   local moment  will  be  exponentially small in  $U/v_f$ such  that exponentially   large  lattices 
will  be  required   to    detect  it.

\begin{acknowledgments}
We would like to  thank  L. Janssen,   Z.Y.  Meng, and  A.  Wipf    for   comments.  
Computational resources were provided by the Gauss Centre for Supercomputing e.V. (www.gauss-centre.eu) through the 
John von Neumann Institute for Computing (NIC)
on the GCS Supercomputer JUWELS~\cite{JUWELS} at J\"ulich Supercomputing Centre (JSC).  
MU  thanks  the  DFG   for financial support  under the projects UL444/2-1. 
FFA   acknowledges financial support from the DFG through the W\"urzburg-Dresden Cluster of Excellence on Complexity and Topology in Quantum 
Matter - \textit{ct.qmat} (EXC 2147, Project No.\ 390858490)   as  well as  the SFB 1170 on Topological and Correlated Electronics at Surfaces and Interfaces (Project No.\  258499086).

\end{acknowledgments}

\bibliography{fassaad, maksim}

\appendix 

\section{Absence of  negative  sign problem} 
\label{appendixA}

 There  are  various  ways   of  formulating  the   QMC.  Although in principle   equivalent,  the    various   formulations   will lead  to  different
 Markov   time  series.    Thereby    the autocorrelation time  for  a   given observable  may  be  formulation  dependent.      The  various  formulations
stem  from   different  ways  writing  the  interaction  term:
 \begin{equation}
      \hat{H}_U  =    - \frac{U}{2}  \sum_{i} \left[
       \left( \hat{a}^{\dagger}_{i}, \hat{b}^{\dagger}_{i} \right)   \sigma_\alpha  \left( \hat{b}^{\phantom\dagger}_{i}, \hat{a}^{\phantom\dagger}_{i} \right)^{T}
      \right]^2.
\end{equation}
In the above,  $\hat{H}_U$  is  independent  on the choice  of  the   Pauli  spin matrix  $\sigma_{\alpha}$.   For the   repulsive  values of  $U > 0 $,    we  can  carry  out  a  HS  decomposition of the perfect  square  term  to  obtain:
\begin{equation}
           Z  \propto   \int D \left\{ \phi(i,\tau)   \right\}     e^{-  S_{\alpha}( \phi(i,\tau)    )   }
\end{equation}
with
\begin{equation}
        S_{\alpha}( \phi(i,\tau)  )   = \int_{0}^{\beta} d \tau \sum_{i} \frac{\phi^{2}(i,\tau)}{2U}        -  \log \text{Tr}  { \cal T} e^{- \int_{0}^{\beta} d  \tau   \hat{h},_{\alpha}(\tau)}
\end{equation}
\begin{equation}
         \hat{h}_{\alpha}(\tau)     =   \hat{H}_0    +    \sum_{i}  \phi(i,\tau) \left( \hat{a}^{\dagger}_{i}, \hat{b}^{\dagger}_{i} \right)   \sigma_\alpha  \left( \hat{b}^{\phantom\dagger}_{i}, \hat{a}^{\phantom\dagger}_{i} \right)^{T},
\end{equation}
and
\begin{equation}
\hat{H}_0 =  -v_F \sum_{i=1} ^{L} \sum_{r=-L/2}^{L/2} t(r) \left(  \hat{a}_{i}^{\dagger} \hat{b}_{i+r}^{} + \hat{b}^{\dagger}_{i+r}  \hat{a}_{i}^{}   \right).
 \end{equation}
We  again stress  that the partition  function is $\alpha$-independent  but that  $\hat{h}_{\alpha}(\tau) $  has  an explicit  $\alpha$-dependency.

Using  the  Majorana  representation of    Eq.~\ref{majorana:eq}  we  obtain  different  expressions  for     various choices of  $\alpha$.
Below  we  go  through them one by one.

\noindent
$ \bullet  $ $\sigma_x$  \\
In this  case,
\begin{equation}
         \hat{h}_x(\tau)  =  v_F \sum_{i=1} ^{L} \sum_{r=-L/2}^{L/2}    \frac{i t(r)}{4}\hat{\ve{\gamma}}_{i}^{T} \tau_x  \hat{\ve{\gamma}}_{i+r}^{}  +
         \sum_{i}     \frac{\phi(i,\tau) }{4}\hat{\ve{\gamma}}_{i}^{T} \tau_y \hat{\ve{\gamma}}_{i}^{}.
\end{equation}
We  see  that   the  operators  $\hat{T}^{+}$,  and  $ \hat{T}^{-}$  with
\begin{equation}
        \hat{T}^{+}  \gamma_i  \hat{T}^{+,-1}     =  T^{+}  \gamma_i  \, \,   \text{  with } T^{+} =  i \tau_y \tau_x \sigma_x K
\end{equation}
and
\begin{equation}
\hat{T}^{-}  \gamma_i  \hat{T}^{-,-1}    =  T^{-} \gamma_i  \, \,   \text{  with } T^{-} = \tau_z i \sigma_y K
\end{equation}
where $K$  denotes  complex  conjugation leaves  $\hat{h}(\tau)$ invariant.     Furthermore
$ \left( T^{\pm} \right)^2  =  \pm 1$  and    $  \left\{ T^{+}, T^{-} \right\}  = 0$.  Hence according to  Ref.~\cite{Li16}     the  Hamiltonian  falls into the
so called  Majorana class  and  does  not  suffer for the negative sign problem.

\noindent
$ \bullet  $ $\sigma_y$  \\
For  this  choice,
 \begin{equation}
         \hat{h}_y(\tau)  =  v_F \sum_{i=1} ^{L} \sum_{r=-L/2}^{L/2}    \frac{i t(r)}{4}\hat{\ve{\gamma}}_{i}^{T} \tau_x  \hat{\ve{\gamma}}_{i+r}^{}  +
         \sum_{i}   \frac{\phi(i,\tau)   }{4}\hat{\ve{\gamma}}_{i}^{T} \sigma_y  \tau_x \hat{\ve{\gamma}}_{i}^{}.
\end{equation}
such  that  we   have  to  choose:
\begin{equation}
        T^{-} =  i \tau_y K  \, \, \text{  and } \, \,  T^{+}  =   \tau_z K
\end{equation}
to  show  that the model is in the Majorana class.

\noindent
$ \bullet  $ $\sigma_z$  \\
For  this  choice,
 \begin{equation}
         \hat{h}_z(\tau)  =  v_F \sum_{i=1} ^{L} \sum_{r=-L/2}^{L/2}    \frac{i t(r)}{4}\hat{\ve{\gamma}}_{i}^{T} \tau_x  \hat{\ve{\gamma}}_{i+r}^{}  -
         \sum_{i}   \frac{\phi(i,\tau)   }{4}\hat{\ve{\gamma}}_{i}^{T} \sigma_y  \tau_z \hat{\ve{\gamma}}_{i}^{}.
\end{equation}
such  that  we   have  to  choose:

\begin{equation}
        T^{-} =  i \tau_y K  \, \, \text{  and } \, \,  T^{+}  =  \tau_z  \sigma_x K
\end{equation}
to  show  that the model is in the Majorana class.

\section{Scaling dimension of $ \hat{S}^z$ operator}
\label{appendixB}

Due to the systemic failure of the SLAC fermion approach in describing the physics of the $ 1+1$ dimensional Helical Luttinger liquid,
which is especially characterized by the spontaneous breaking of $U(1)$ symmetry, we cannot expect the scaling behavior of the bosonization results of Eq.~\ref{correlations.eq}   to hold. Taking the $\hat{S}^z$ operator as an example, its equal-time structure factor shows a sharp (or smooth) cusp around   $k=0$ ($k=\pi$) as the interaction strength $U$ increases, as depicted in Fig~\ref{fig:C}(h). This indicates a violation of the $1/r^2$ scaling relation based on naive bosonization.

Generally the (equal-time) real space correlation function of $\hat{S}^z$ operator is:
\begin{equation}
  C_{ \hat{S}^z }(r) = a r^{ - 2\Delta_0^{\hat{S}^z} }
  +  b (-1)^r r^{ -2\Delta_\pi^{ \hat{S}^z } }
\end{equation}
where $\Delta_0^{\hat{S}^z} $ and $\Delta_\pi^{\hat{S}^z}$ are the scaling dimensions
at $k=0$ and $k=\pi$,  and   $a$ and $b$ here are non-universal constants.
We use the quantity $B_{\hat{S}^z}(k, L/2)$, as defined in Eq.~\ref{SF}, to extract the scaling dimension.  In  particular  in the
$L \rightarrow  \infty   $ limit  we    expect:
\begin{equation}
 \begin{aligned}\label{eq:scal_b}
& B_{ \hat{S}^z }( k=0, L/2) \propto L^{ - 2 \Delta_0^{ \hat{S}^z } + 1 }   \\
& B_{ \hat{S}^z }(k=\pi, L/2) \propto L^{ - 2 \Delta_\pi^{ \hat{S}^z } + 1 }
 \end{aligned}
\end{equation}
This approach for determining the scaling dimension relies on calculating the difference between the left and right derivatives of the structure factor. Therefore, it provides correct  results within the range of $0.5 < \Delta \leq 1$. For $\Delta > 1$, the structure factor becomes a smooth function at the selected momentum point, making it difficult for our approach to distinguish between an exponential or power-law decay.

\begin{figure}
\centering
\includegraphics[width=0.45\textwidth,height=0.9\textheight]{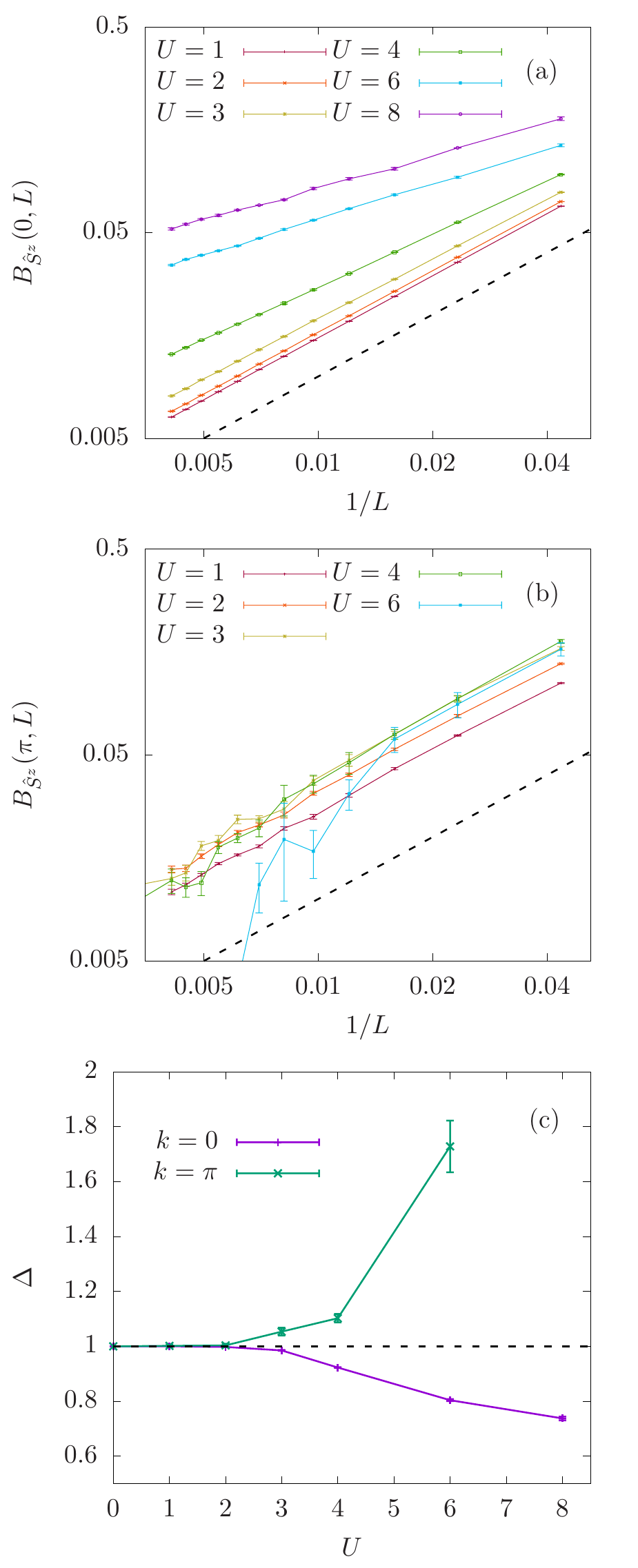}
\caption{
 $1/L$ dependence of $ B_{ \hat{S}^z }( k=0, L/2)$ (a) and  $ B_{ \hat{S}^z }( k=\pi, L/2)$ (b).
 The dashed line is $f(L) = 1/L$, as a guide to the eye.
 The extrapolated scaling dimension $\Delta_0$ and $\Delta_\pi$ as a function of $U$ is ploted in (c).
 The dashed line is $\Delta=1$, as a guide to the eye.
}\label{fig:scal_Z}
\end{figure}

Fig.~\ref{fig:scal_Z}(a) and (b) show that $B_{\hat{S}^z}(k=0, L/2)$ scales linearly as a function of $1/L$, and its slope decreases in the large $U$ case, while $B_{\hat{S}^z}(k=\pi, L/2)$ behaves oppositely. The extrapolated scaling dimension, obtained by fitting the power law function of Eq~\ref{eq:scal_b}, is 1 within the error bars for $U < 3$. For $U > 3$, the values of $\Delta_{\hat{S}^z}$ systematically decrease (increase) for $k=0$ ($k=\pi$), as shown in Fig.~\ref{fig:scal_Z}(c).

To verify the consistency of our results, we also conducted a Density Matrix Renormalization Group (DMRG) simulation of the one-dimensional XXZ chain with long-range interaction, which corresponds to the perturbed Hamiltonian of the SLAC system in the strong coupling limit. Eq.~\ref{Strong.eq} in the main text can be reformulated as:
\begin{equation}
\label{XXZ}
     \hat{H}_{XXZ}  = J \sum_{i,r \in \text{OBC}} \frac{1}{r^2}   \left[ \hat{S}^{z}_{i}  \hat{S}^{z}_{i +r}   -   \hat{S}^{x}_{i}  \hat{S}^{x}_{i +r}    - \hat{S}^{y}_{i}  \hat{S}^{y}_{i +r}    \right].
\end{equation}
where  $J$ is the only energy scale of the Hamiltonian and we set it  to  unity.
Instead of the periodic boundary condition that is considered for SLAC Hamiltonian, we consider an open boundary case here, such that
the long range interaction  in Eq.~\ref{XXZ} is truncated at the boundary.
We implemented the DMRG algorithm in the ITENSOR library \cite{itensor}.
The power law nature of interaction in this system does not lead to a dramatic increase of entanglement in DMRG simulation, and we checked  convergence for bond dimensions up to $\chi=400$.

\begin{figure}
\centering
\includegraphics[width=0.5\textwidth,height=0.9\textheight]{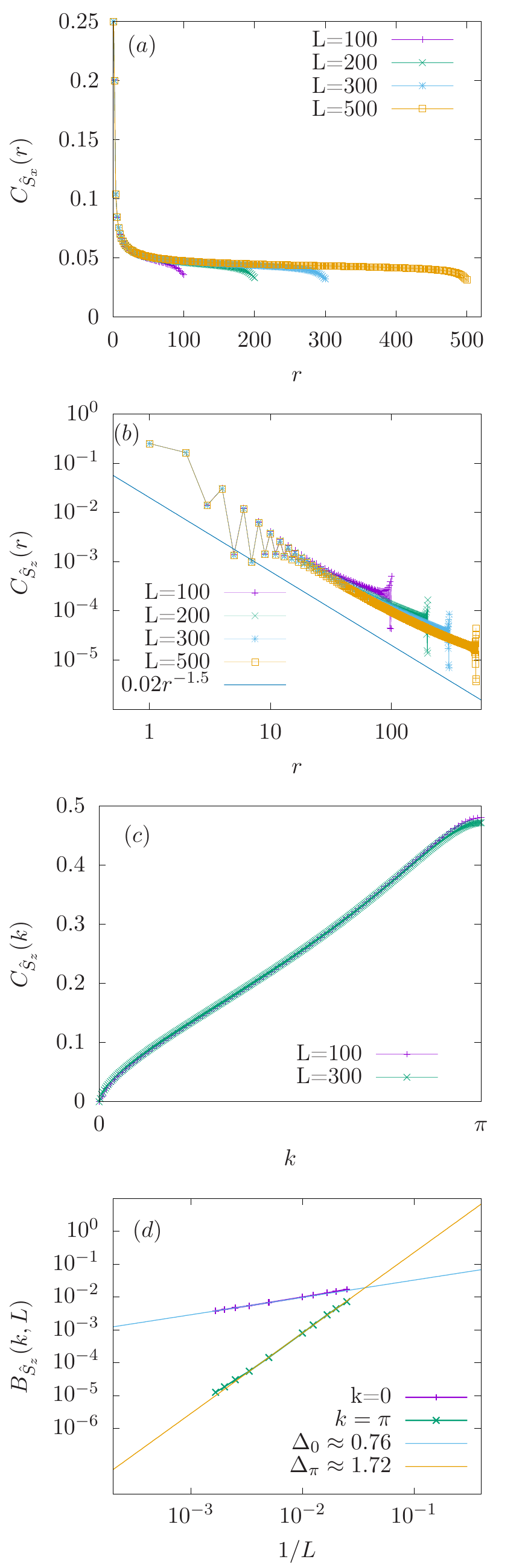}
\caption{
 Ground state property of the XXZ chain based on DMRG simulation.
(a). Real space correlation function of $\hat{S}_x$ operator originated from the left boundary of the chain.  (b). Same as (a), for $\hat{S}_z$ operator. The blue line plots function $0.02r^{-1.5}$, is a guide to the eye.
 (c). Structure factor of the $\hat{S}_z$ operator.
 (d).   $1/L$ dependence of $ B_{ \hat{S}_z }(k, L/2)  $ for $k=0$ and $\pi$ in a logarithmic scale, as well as the two fitted function based on  Eq.~\ref{eq:scal_b} as guides to the eye.
}
\label{fig:XXZ}
\end{figure}

Fig.~\ref{fig:XXZ}(a) displays the long-range correlation of the $\hat{S}_x$($\hat{S}_y$) operator, which is consistent with our numerical results of the SLAC Hamiltonian in the large $U$ regime. On the other hand, the real-space decay of the $\hat{S}_z$ operator shows an algebraic scaling behavior, as shown in Fig.~\ref{fig:XXZ}(b) and (c). It should be noted that the seemingly square root behavior of $C_{ \hat{S}_z }(k) $ at $k \approx 0$ also fits well with the plot of the SLAC system in the large $U$ limit, as depicted in Fig.~\ref{fig:C}(h).

Finally, we also performed a scaling analysis for $ B_{ \hat{S}^z }( k, L/2)$
base on Eq.~\ref{eq:scal_b}.
As shown in Fig.~\ref{fig:XXZ}(d), $B_{ \hat{S}^z }( k, L/2)$  displays a
nice power law behavior as function of $1/L$.  A collective fit based on Eq.~\ref{eq:scal_b}, using system sizes of $L=200,300,400,500,600,800$ and $1000$,
gives the scaling dimension of $ \hat{S}_z $ operator at $k=0$.
\begin{equation}
  \Delta_0^{ \hat{S}^z } =  0.762(2)
\end{equation}
Note  that  at  $k=\pi$ the  $  \Delta_\pi^{ \hat{S}^z } > 1  $   such  that  the  structure  factor  at   $k=\pi$   is  a  smooth  function    consistent  with
an  exponential  decay of  staggered  fluctuations.

\section{Spin wave analysis  of  the  $1/r^2$ XXZ  chain}
\label{appendixC}

\begin{figure}
\centering
\includegraphics[width=0.45\textwidth]{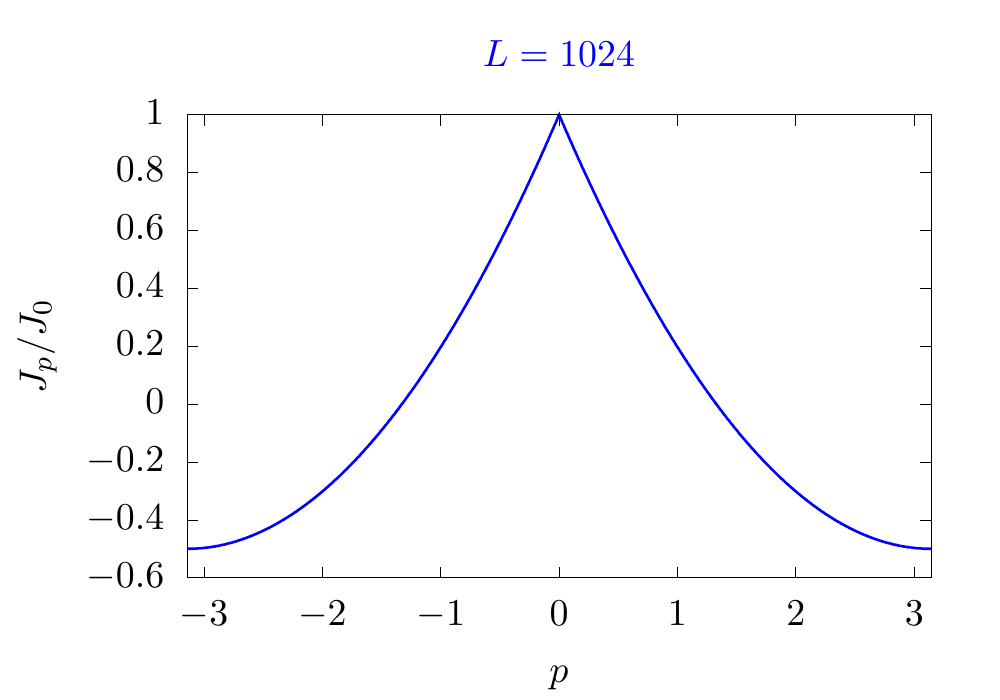}
\caption{
$J_p /J_0 $  for  the  SLAC  hopping  of  Eq.~\ref{Slac_t.eq}.
}\label{fig:spin_wave}
\end{figure}

In this appendix,  we  carry  out a  spin-wave analysis of the  XXZ  model   with $1/r^2$   exchange     derived  in Eq.~\ref{Strong.eq}.   We  will see  that  the  long ranged interaction stabilizes  the  order  and  that  the  scaling  dimension of  the  $\hat{S}^z$  operator  reproduces  the   DMRG  results.     We  first  carry  out  a  canonical  transformation:
\begin{equation}
        \hat{\tilde{\ve{S}}}_{i}     =  R(\ve{e}_y, \pi/2)   \hat{\ve{S}}_{i}
\end{equation}
where  $  R(\ve{e}_y, \pi/2)   $  is an  SO(3)    rotation   around  the axis $\ve{e}_y$  and  with  angle  $\pi/2$  such  that  e.g.  $\hat{\tilde{S}}^z_{i}      =  \hat{S}^x_i$.   Hence, Eq.~\ref{Strong.eq}    maps  onto
\begin{equation}
\label{Strong1.eq}
     \hat{H}_{XXZ}  =  \sum_{i,r=-L/2}^{L/2}   J(r) \left[ \tilde{\hat{S}}^{x}_{i}  \tilde{\hat{S}}^{x}_{i +r}   -
      \hat{\tilde{S}}^{y}_{i}  \hat{\tilde{S}}^{y}_{i +r}    - \hat{\tilde{S}}^{z}_{i}  \hat{\tilde{S}}^{z}_{i +r}    \right].
\end{equation}
with  $J(r)  =    \frac{4v_f^2}{U} t^2(r)  $.     We  will  assume  long-ranged    ferromagnetic  magnetic  order   along   the $\hat{\tilde{S}}^{z}_{i}  $    quantization axis  and  adopt   the  Holstein-Primakov  representation:
\begin{eqnarray}
        & & \hat{\tilde{S}}^{z}_{i}   = -  \hat{b}^{\dagger}_i  \hat{b}^{\phantom\dagger}_i   + S    \nonumber  \\
        & & \hat{\tilde{S}}^{+}_{i}   =   \sqrt{  2S - \hat{b}^{\dagger}_i  \hat{b}^{\phantom\dagger}_i  }  \, \,   \hat{b}_i
\end{eqnarray}
with  $  \left[ \hat{b}^{}_i,  \hat{b}^{\dagger}_j  \right] = \delta_{i,j} $.       For  small  fluctuations around  the ordered state,   $ \left<  \hat{b}^{\dagger}_i  \hat{b}^{\phantom\dagger}_i \right> \ll  2S$  we     obtain:
\begin{eqnarray}
        \hat{H}_{XXZ}  = & &  E_{MF}  + S\sum_{p} \left(  2J_0 \hat{b}^{\dagger}_p \hat{b}^{\phantom \dagger}_p   + J_p \hat{b}^{\dagger}_{-p} \hat{b}^{\dagger}_p    +   J_{-p}  \hat{b}^{\phantom\dagger}_{-p} \hat{b}^{\phantom\dagger}_p  \right)   \nonumber  \\
            & & +  {\cal O} \left(S^0 \right).
\end{eqnarray}
The  first  term  corresponds  to the  Weiss mean-field   energy,   $ E_{MF}   =   - \sum_{i,r} J(r) S^2 $    and  scales as  $S^2$.
The  second   term scales  as  $S$ and    describes  spin wave
fluctuations    with  $  J_p  =   \sum_{r} e^{ip r} J(r) $  and  $ \hat{b}_p   =  \frac{1}{\sqrt{L}}  \sum_{i} e^{ip r } \hat{b}_i $.       Lower order in  $S$ are
neglected.
We  diagonalize  the above Hamiltonian  with the Bogoliubov  transformation:
\begin{equation}
          \hat{a}^{\dagger}_p  =   \cosh \left(   \theta_p \right)  \hat{b}^{\dagger}_p    + \sinh \left(   \theta_p \right)  \hat{b}^{\phantom\dagger}_{-p}
\end{equation}
and
\begin{equation}
        \tanh(2 \theta_p)   =   \frac{J_p}{J_0}
\end{equation}
to  obtain
\begin{equation}
           \hat{H}_{XXZ}     =    2S J_0  \sum_{p}   \sqrt{1 -  \frac{J_p^2}{J_0^2}}  \, \,   \hat{a}^{\dagger}_p   \hat{a}^{\phantom\dagger}_p    +  C
\end{equation}
up  to  a  constant $C$.
For  the  SLAC  hopping  of  Eq.~\ref{Slac_t.eq},  $\frac{J_p}{J_0} $  is  plotted  in Fig.~\ref{fig:spin_wave}.  As  apparent,  in the vicinity of     zero  momentum it  scales  as
\begin{equation}
         \frac{J_p}{J_0}    \simeq   1  -    \alpha  | p |.
\end{equation}
As   a  consequence,
\begin{equation}
         \frac{1}{L} \sum_p \langle  \hat{b}^{\dagger}_{p} \hat{b}^{\phantom\dagger}_p   \rangle    =   \frac{1}{2\pi} \int_{-\pi}^{\pi}   d p  \left(   \frac{1}{\sqrt{1- \left( \gamma_p/\gamma_0\right)^2 }  } -1  \right).
\end{equation}
In the vicinity of  zero momentum, the   integral takes the form,  $\int_0^{\Lambda}   d p  \frac{1}{\sqrt{|p|}}$ ,    and  converges.  Thereby fluctuations    around  the mean-field solution remain  small   and   the   spin-wave approximation  is  justified.  Note  that for   short  ranged
hopping,  $  \frac{J_p}{J_0}    \simeq   1 - \alpha p^2  $   such  that    $ \frac{1}{L} \langle  \hat{b}^{\dagger}_{p} \hat{b}^{\phantom\dagger}_p   \rangle $   diverges  and  fluctuations  destroy  the order.

Finally   we  compute  the    transverse spin-spin  correlations:
\begin{eqnarray}
        \langle  \hat{S}^{z}_{r} \hat{S}^z_{0}   \rangle     & & = \frac{S}{ 8 \pi  }  \int_{-\pi}^{\pi}   d p e^{i p  r}  \frac{1 - J_p/J_0}{\sqrt{1-\left(J_p/J_0\right)^2}}          \nonumber  \\
  & &  \propto   \frac{S}{ 8 \pi  }  \int_{-\pi}^{\pi}      d p e^{i p  r}  \sqrt{\alpha |p| }    \propto   \frac{1}{r^{1.5}}.
\end{eqnarray}
This  provides  a  very  good   match  with the DMRG result.

\end{document}